\documentstyle[12pt,aaspp]{article}
\begin{document}

\title{Integrated UV Spectra and Line Indices of M31\\
Globular Clusters and the Cores of Elliptical Galaxies\altaffilmark{1,2}} 

\author{Jerry M. Ponder\altaffilmark{3}, David Burstein\altaffilmark{3},}
\author{Robert W. O'Connell\altaffilmark{4},}
\author{James A. Rose\altaffilmark{5},}
\author{Jay A. Frogel\altaffilmark{6}, Chi--Chao Wu\altaffilmark{7},}
\author{D. Michael Crenshaw\altaffilmark{8},}
\author{Marcia J. Rieke\altaffilmark{9}, and Michael Tripicco\altaffilmark{10}}

\altaffiltext{1}{Based on observations with the NASA/ESA Hubble Space
Telescope, obtained at the Space Telescope Science
Institute, which is operated by the Association of Universities for Research
in Astronomy, Inc., under NASA contract NAS 5-26555.}
\altaffiltext{2}{Based partly on observations obtained with the
{\it International Ultraviolet Explorer Satellite}, which was sponsored and
operated by the United States National Aeronautics and Space Administration,
by the Science Research Council of the United Kingdom, and by the European
Space Agency.}
\altaffiltext{3}{Department of Physics and Astronomy, Arizona State University.
Box 871504, Tempe, AZ 85728-1504}
\altaffiltext{4}{Astronomy Department, University of Virginia, P.O. Box 3818,
Charlotteville, VA 22903-0818}
\altaffiltext{5}{Department of Physics and Astronomy, 
University of North Carolina, CB 3255, Phillips Hall, Chapel Hill, NC 
27599-3255}
\altaffiltext{6}{Department of Astronomy, The Ohio State University, 174 W.
18th Ave, Columbus, OH 43210}
\altaffiltext{7}{Science Programs, Computer Sciences Corporation, 
Space Telescope Science Institute, 3700 San Martin Dr., Baltimore, MD 21218}
\altaffiltext{8}{Catholic University of America, Department of Physics, 
Washington, D.C. 20064}
\altaffiltext{9}{Steward Observatory, University of Arizona, Tucson, AZ
85721}
\altaffiltext{10}{NASA/Goddard Space Flight Center, Code 664, Greenbelt, MD
20771}

\begin{abstract}

We present observations of the integrated light of four M31 globular clusters
(MIV, MII, K280, and K58) and of the cores of six elliptical galaxies (NGC
3605, 3608, 5018, 5831, 6127, 7619) made with the Faint Object Spectrograph
(FOS) on HST.  The spectra cover the range $\rm 2200 \AA$ to $\rm 4800 \AA$ at
a resolution of $\rm 8 \AA$ with S/N ratio $>20$ and flux accuracy of $\sim
5$\%.  To these data we add from the literature IUE observations of the dwarf
elliptical M32, Galactic globular clusters and Galactic stars. The stellar
populations in these systems are analyzed with the aid of mid--UV and near--UV
colors and absorption line strengths.  Included in the measured indices is the
key NH feature at 3360$\rm \AA$.  We compare these line index measures to the
2600--3000 colors of these stars and stellar populations.

We find that the M31 globular clusters, Galactic globular clusters/Galactic
stars, and elliptical galaxies represent three distinct stellar populations,
based on their behavior in color-line strength correlations involving Mg II,
NH, CN, and several UV metallic blends. In particular, the M31 globular
cluster MIV, as metal-poor as the Galactic globular M92, shows a strong NH
3360 feature.  Other line indices, including the 3096$\rm \AA$ blend that is
dominated by lines of Mg I and Al I, show intrinsic differences as well. We
also find that the broad-band line indices often employed to measure stellar
population differences in faint objects, such as the 4000$\rm \AA$ and the Mg
2800 breaks, are disappointingly insensitive to these stellar population
differences.

We find that the hot ($T > 20000$K) stellar component responsible for the
``UV-upturn'' at shorter wavelengths can have an important influence on the
mid-UV spectral range (2400--3200 $\rm \AA$) as well. The hot component can
contribute over 50\% of the flux at 2600 $\rm \AA$ in some cases and affects
both continuum colors and line strengths. Mid-UV spectra of galaxies must be
corrected for this effect before they can be used as age and abundance
diagnostics.

Of the three stellar populations studied here, M31 globular clusters and 
elliptical galaxies are more similar to each other than each is to the 
Galactic stellar populations defined by globular clusters and nearby stars.
Similarities between the abundance pattern differences currently identified
among these stellar populations and those among globular cluster stars 
(N,Al enhancements) present a curious coincidence that deserves future 
investigation.

\end{abstract}


\section{Introduction}

     The integrated spectra of galaxies are notoriously difficult to interpret
in terms of their overall metallicity and age (cf. the excellent review by
\cite{CWB96}). Every part of the spectral energy distribution (SED) of a
stellar system yields a different clue to a complex puzzle.  In the
space-ultraviolet part of these SEDs we can unambiguously study the 
contribution of the hottest stars in stellar populations.  The far-UV ($\rm
\lambda \lambda 1250 - 2000 \AA$) and mid-UV ($\rm \lambda \lambda 2000-3300
\AA$) can most easily be studied in galaxies at high--redshift.  Certainly, a
prerequisite towards understanding the UV SEDs of galaxies at high redshift is
to understand what we see in the UV SEDs of nearby stellar systems.

     Towards the goal of understanding the UV SEDs of stellar populations,
several of us have used the International Ultraviolet Explorer (IUE) to
assemble a medium resolution (6$\rm \AA$) far-UV and mid-UV library
of the spectral energy distributions of stars (\cite{FOBW90}; hereafter FOBW;
\cite{FOBW92}; \cite{lietal98}).  In Fanellli et al. (1990, 1992) we found
that the mid-UV color and absorption line indices have dependences on
temperature and metallicity which are strong and distinct from those derived
from spectra longward of 3300$\rm \AA$. In particular, Fanelli et al. (1992)
found that absorption line measures using the usual index method (cf.
\cite{O73}; \cite{F73}) yield mid-UV indices which are primarily sensitive to
stellar temperature but insensitive to metallicity.  One index in particular
(MgII at 2800$\rm \AA$) is found to be {\it inversely} metallicity-sensitive
for F--G dwarf stars, in that it gets weaker with increasing metallicity (cf.
FOBW; \cite{smithetal}). In contrast, the mid-UV colors, which measure
mean line blanketing, are found to the most sensitive to metallicity
differences among stellar populations (FOBW).

     Several groups have studied the far-UV and mid-UV SEDs of early--type
galaxies and spiral galaxy bulges (e.g., \cite{BBBFL88}; \cite{BFD95}), as
well as of M31 globular clusters (e.g. \cite{CCBFK}; \cite{CB88};
\cite{CKCF90}). From these studies we have learned that early-type galaxies
contain a hot stellar population (T $>$ 20000K) whose amplitude, relative to
the optical, is well correlated with stellar population absorption line
strengths.  Various studies indicate that this hot stellar population is
important in galaxy spectra below $\rm \sim 3200 \AA$; consists mainly of hot,
metal rich, extreme horizontal branch stars and their post-HB descendents;
and is very sensitive to the age and abundance of their parent population
(e.g. \cite{GR90}, \cite{BFD95}, \cite{dorman95}, \cite{TCBF96}, and
\cite{yi97}).

     The integrated light of Galactic globular clusters can usually be
interpreted in terms of combinations of spectra of well understood field star
populations near the Sun (cf. \cite{christensen}; \cite{rose94};
\cite{rosedeng98}).  However, integrated optical line indices of M31 globular
clusters exhibit anomalies which seem to distinguish them both from local
field stars and Galactic globular clusters (cf. \cite{vdB69}; \cite{BFGK84};
\cite{trip89}; \cite{BH90}), especially concerning CNO-related indices (such
as CN).

    In the present paper we report on Hubble Space Telescope/Faint Object
Spectrograph (HST/FOS) observations with small ($\le 1\arcsec$) apertures of
the $\rm 2300 - 4800 \AA$ SEDs of four M31 globular clusters and six bright
elliptical galaxies. The galaxies chosen have old stellar populations that span
the range known from optical observations (cf. \cite{FWBDDLBT},
\cite{gonzalez}). The M31 globular clusters were chosen to span a range of CN
line strengths; all four have had their color-magnitude diagrams determined by
HST imaging (cf. \cite{fusipec96}). We combine these data with IUE-obtained
mid-UV spectra of Galactic globular clusters and the dwarf elliptical galaxy
M32. In \S 2 we discuss the FOS observations and present the observed spectra.
In \S 3 we present a summary of the Galactic globular clusters and the M32 IUE
data, which are detailed elsewhere. \S 4 details how the relationship between
mid-UV line indices and the 2600-3000 color compare for the galaxies,
M31 globular clusters, IUE data for Galactic stars and IUE data for
Galactic globular clusters. \S 5 makes the analogous comparisons for the
near--UV line indices.  The various differences and similarities seen among
the line stength-color relations for these stellar populations are discussed
in \S 6.  The main results of this paper are summarized in \S 7, and an
Appendix discusses how the HST/FOS acquired each object for its science
aperture.

\section{Observations and Reductions}

\subsection{Target Selection}

Of the four M31 globular clusters chosen for study here, three (MII = K1,
MIV = K219, and K280) had been previously studied in integrated light by 
Burstein et al. (1984) while a fourth (K58) comes from the integrated light
study of Brodie \& Huchra (1990). [The ``K'' names of the globular clusters
are taken from the listing of \cite{SKHV77}; ``M'' names from \cite{ME53};
``Bol'' names from \cite{batt80} and \cite{batt87}.] 
These globular clusters span the range from the most metal-poor (MIV) to the
most metal-rich (K58) M31 clusters known.

The six galaxies were chosen to span a wide range of known stellar populations
among old stellar systems as inferred by Gonzalez (1993). Gonzalez bases his
estimates of galaxy ages on analyzing the strengths of $\rm H\beta$, mean Fe
indices and the optical $\rm Mg_2$ index with the galaxy population models of
Worthey (1992).  (Here, we explicitly state ``the optical $\rm Mg_2$ index,''
so as to distinguish from the UV-based Mg-indices.)  In his reanalysis of
these data using the more up-to-date models of Charlot et al. (1996), Trager
(1996) finds the rank-ordering of galaxy ages to be the same as Gonzalez.  We
note that while both Gonzalez and Trager provide independent estimates of age
and metalliticy for their galaxies, in our samples age and metallicity (at
least, strong absorption lines) are correlated (stronger lines $\propto$ 
older ages).

The galaxies (in order of decreasing optical Mg$_2$ strength) are: NGC~7619
(Mg$_2$ = 0.358), NGC~3608 (0.329), NGC~6127 (0.322), NGC~5831 (0.303),
NGC~3605 (0.241), NGC~5018 (0.210) and M32 (0.198). NGC~5018 is added to this
sample, although not specifically observed by Gonzalez, as it has UV/optical
colors and optical line strengths similar to that of M32, yet has the
luminosity of a giant elliptical (\cite{BBB93}).  Of this sample, only M31 and
M32 have far--UV SEDs available (both from IUE; \cite{BBBFL88}). On the scale
defined by the Worthey models, ages for these objects range from above 12 Gyr
at the oldest, to 3--5 Gyr at the youngest, or Mg$_2$ = 0.35 at the
strongest-lined and 0.20 at the weakest-lined. We also note that none of these
objects is expected to have a significant stellar population component of age
less than 1 Gyr (cf. \cite{gonzalez}; \cite{Trager1}).

\subsection{Observations}

FOS observations for the globular clusters were originally scheduled under
HST/GO program P2298 for Cycle 1, but were obtained in Cycles 1, 2 and 4.  FOS
observations for the galaxies were taken under HST/GO program P6585 in Cycle~6.
Table~1 contains the log of observations, with the HST archival name for each
spectrum given for easy reference.  Also given in Table~1 are J2000 positions
for the objects observed, as well as their optical Mg$_2$ values and
the radial velocities we used for them. The Mg$_2$
measures come from Burstein et al. (1984) for MII, MIV and K280. The Mg$_2$
values for K58 is transformed from that listed by Brodie \& Huchra (1990) (by
applying a net offset of +0.026 mag to bring their measures in accord with
those of Burstein et al.).  The Mg$_2$ measures for the galaxies come from
Trager et al. (1998). All Lick Mg$_2$ observations were observed with a
1.4$''$ $\times$ 4$''$ slit (cf. \cite{Trager2}), so sample the clusters and
galaxies at angular resolutions comparable to the HST observations.
The radial velocities for the M31 globular clusters come from Huchra, Brodie 
\& Kent (1991), while those for the galaxies come from de Vaucouleurs et al.
(1991).

Given the diffuse nature of the objects studied, the FOS aperture for each
observation had to be centered via the FOS ACQ/PEAK-UP procedure to obtain
the required positional accuracy within the nominal 1$''$ circular aperture
(for Cycles 1 and 2, pre-COSTAR, which becomes 0.86$''$ for Cycles 4 and 6,
post-COSTAR).  The manner and accuracy of target acquisition are detailed
in Appendix~A.

The M31 globular clusters were observed with three FOS gratings:  G130H, G270H
and G400H.  The G130H grating was used, despite its limited wavelength
coverage, as it was suggested by previous IUE observations (e.g. \cite{CCBFK})
that some of the M31 globular clusters might contain a substantial hot stellar
population.  Although we took long integrations with the G130H grating, no
significant signal in the full 1300--1600$\rm \AA$ bandpass was measured for
any of the M31 clusters (see discussion below).  Thus, we do not find any very
hot stellar populations in these clusters that falls within our aperture (of
diameter 2.9 pc at the distance of M31).  Combination of the G270H plus G400H
gratings give us continuous wavelength coverage from 2300$\rm \AA$ to 4800$\rm
\AA$, with essentially no observational overhead.

As the galaxy spectra were obtained after the M31 globular cluster spectra, we
knew that we would not likely get any scientific information from the G130H
grating in a reasonable number (i.e., $< 20$) of orbits, so it was decided to
concentrate only on getting accurate spectra with the G270H and G400H
gratings.  It typically took 6 orbits to obtain the spectra we show here for
the galaxies, including 3.5 orbits to do the ACQ/PK procedure.  Obtaining
accurate UV spectra of diffuse objects with HST/FOS was expensive in time.

\subsection{Data Reduction}

Our data reduction procedures followed those recommended by the HST/FOS Data
Products Guide (\cite{DPG}; hereafter DPG).  The data sent to us were
processed by Space Telescope Science Institute with the Routine Science Data
Processing software, which produces files of fluxes, wavelengths, errors, data
qualities as well as raw data. As a check, we ran the raw data through our
local IRAF/STSDAS programs and found no differences between the HST-provided
data and our re-reduced data.

Flux and wavelength files were combined as recommended by the DPG, including
resampling and Hanning smoothing of the data.  Since the observed spectral
resolution is finer than the spectral resolution permitted by the nominal
1$''$ aperture, the spectra were further smoothed by a boxcar window of
sufficient size to produce spectral resolution of $\rm 5.5 \AA$ for the G270H
grating and $\rm 8.8 \AA$ for the G400H grating. The G270H spectra of two
galaxies (NGC~6127, NGC~5018) are noisier than for the other objects in this
sample.  Examination of the error spectrum for NGC~6127 shows that the
apparent "emission" line at the position of the Mg 2800 line is an artifact
resulting from low signal-to-noise and a diode in the FOS that had just gone
bad.

As stated previously, we only were able to obtain upper limits for the
observations of the M31 globular clusters obtained with the FOS G130H grating.
Integrating the flux observed with this grating from 1300 to 1600$\rm \AA$ we
obtain: $\rm 9.3 \pm 19 \times 10^{-17}$ ergs s$^{-1}$ cm$^{-2}$ for MII,
$\rm 0.8 \pm 19 \times 10^{-17}$ ergs s$^{-1}$ cm$^{-2}$ for K58 and $\rm 1.3
\pm 18 \times 10^{-17}$ ergs s$^{-1}$ cm$^{-2}$ for MIV.  The FOS shutter did
not open for the observation of K280, so we use this observation to give us
the background rate.

All spectra are dereddened, using the values of E(B--V) given in Table~1 and
the reddening law of Cardelli et al. (1989), updated by O'Donnell (1994) for
the near-UV. E(B--V) values are taken from Burstein \& Heiles (1984) for the
galaxies and for M31 (assuming the clusters are reddened only by foreground
Milky Way extinction). Spectra of all objects were then Doppler-shifted to
their rest wavelengths using the radial velocities given in Table~1. The
de-reddened fluxes are given at rest wavelengths, and will be made available
for each object in our sample in ASCII format via various electronic means,
including: anonymous ftp; electronic preprint, and, eventually, the
Astronomical Data Center.

The resulting spectra are shown in linear flux units in Figures~1a and 1b for
the globular clusters and in Figures 2a and 2b for the galaxies, at their rest
wavelengths.  Figures 3, 4a and 4b show the full HST/FOS spectra of the
clusters and the galaxies (plus the IUE mid-UV spectrum of M32) in log flux
units.  Spectra from the G270H grating and the G400 grating are shown
separately in parts a and b of Figures 1 and 2 for clarity.  A test of the
zero point accuracy of the relative fluxes obtained from the two spectra can
be derived from the 70$\rm \AA$ overlap in the region 3200 -- 3270 $\rm \AA$.
As shown in Figures 3 and 4, the relative fluxes agree within the nominal
value of 5\% we are led to expect by the FOS Handbook.

\subsection{Line Indices}
	
Our spectra cover a range of UV and optical wavelengths for which no single
unified set of absorption line and continuum indices has yet been defined.
Rather, our choice of indices and colors come from three separate studies: a)
The range 2300 to 3100 $\rm \AA$ has the mid-UV indices defined by FOBW. b)
The range 3100 to 3800 $\rm \AA$ have several indices defined by Davidge \&
Clark (1994), including the very useful NH molecular band at 3360$\rm \AA$. c)
The range 3800 to 4800 $\rm \AA$ is covered by Lick Observatory photometric
systems (\cite{worthey94}).  A list of the definitions of the line strength
indices measured for our spectra, and the source of each index, is given in
Table~2.

To determine the index values, we use the definitions of spectral
indices given by Gonzalez (1993):

\begin{equation}
\rm I_{a} = \int_{\lambda_{c_{1}}}^{\lambda_{c_{2}}}
(1-\frac{S(\lambda)}{C(\lambda)})\ d\lambda
\end{equation}

\begin{equation}
\rm I_{m} = -2.5 log_{10} {\int_{\lambda_{c_{1}}}^{\lambda_{c_{2}}}
\frac{S(\lambda)}{C(\lambda)} \ d\lambda}
\end{equation}

\noindent where $S(\lambda)$ represents the object flux at each wavelength
within the central bandpass, $C(\lambda)$ represents the pseudo-continuum flux
for $S(\lambda)$ within the central bandpass, and $\lambda_{c_{1}}$ and
$\lambda_{c_{2}}$ are the limits of the central bandpass. All fluxes in this
paper are computed per unit wavelength.  Equivalent width indices computed
from Equation 1 are measured in angstroms; indices computed from Equation 2
are measured in magnitudes. Equation 2 is essentially identical to the line
index definition of FOBW, differing negligibly ($<<0.001$ mag) in practice. 
The pseudo-continuum within the central bandpass is determined by linear
interpolation between the red and blue pseudo-continuum sidebands defined for
each absorption feature. The wavelength ranges of the red and blue sidebands
and the central bandpasses for the absorption features measured in this paper
are given in Table~2. The spectral break indices and the intermediate band
color index 2600-3000 are defined as the ratio of the mean fluxes within the
appropriate bands converted to magnitudes.  To account for uncertainties in
the radial velocites and in the wavelength scales used, each spectral line
measurement was adjusted interactively.  We adjusted the continuum and feature
passband positions (typically by less than 2 pixels) by comparing them with
their placement given in the papers that define these line indices. The
spectral indices and colors for all observations are given in Tables~3a,b.

Errors were determined from the error spectra provided by the pipeline
reduction process.  These error spectra are computed by propagating the errors
at each point with the assumption that the dominant noise in the raw data is
photon noise.  We then determine the mean error based on photon statistics
($\rm \delta S(\lambda)$ or $\rm \delta C(\lambda)$ for each bandpass and
propagate those errors through the previous equations, with the results:

\begin{equation}
\rm \frac{\delta I_a}{I_a} = \frac{(\lambda_1 - \lambda_2)}{C(\lambda)}
\sqrt{[(\delta
S(\lambda))^2 + {(\frac{\delta C(\lambda) \times S(\lambda)}{C(\lambda)})}^2
} \, ,
\end{equation}

\begin{equation}
\rm \delta I_{mag} = \frac{2.5}{ln(10)} \sqrt{[\frac{\delta
S(\lambda)}{S(\lambda)}]^{2} + [\frac{\delta
C(\lambda)}{C(\lambda)}]^{2}} \, , and
\end{equation}

\begin{equation}
\rm \delta Break = \frac{2.5}{ln(10)} \sqrt{[\frac
{\delta S(\lambda)}{S(\lambda)}]^{2}} \, .
\end{equation}

\noindent The errors for our measured indices are given in Table~3 in terms
of either magnitudes (for broad features and breaks) or equivalent widths
(in $\rm \AA$). 

\section{Supplemental IUE Data}

To supplement the HST observations, we have also included IUE-obtained data
for the galaxy M32 (from \cite{Buson90}) and for 24 Galactic globular
clusters.  The IUE globular cluster data come from the study of Rose \& Deng
(1998). Rose and Deng surveyed the IUE database and identified 24 Galactic
globular clusters which have been observed by IUE in their core regions, and
which have good quality mid-UV IUE low-resolution spectra.  In the case of
M32, we have derived all of the mid-UV line indices and 2600--3000 color
(listed in Table~3a) from the data of Buson et al. In the case of the Galactic
globular clusters, Rose and Deng derived only four of the mid-UV line indices
(Mg I, Mg II, the 2609/2660 and 2828/2921 line breaks) plus the 2600--3000
color.  The data for M32 appear in a subset of the line strength-color
relations we compare to those of stars, while the Galactic globular cluster
data appear only when we intercompare the line strengths of galaxies, M31
globular clusters and Galactic globular clusters.  The IUE obervations act as
a good benchmark against which to differentially compare the HST/FOS
observations of M31 globular clusters and galaxies.

Details of the data reduction of IUE data are given in Rose \& Deng (1998). In
summary, most of the Galactic clusters have only a single IUE observation, but
for the ones with multiple observations, the coadded spectra were used. We
list the 24 globular clusters in Table~4, together with their IUE-based line
strength measures from Rose \& Deng, estimates of [Fe/H] taken from Harris
(1997), and $\rm Mg_2$ measurements taken from Burstein et al. (1984). As the
IUE aperture of $10\arcsec \times 20\arcsec$ covers a fairly large region of
the cores of most clusters, we believe the spectra are representative of the
integrated mid-UV light of these cores.  All of the IUE spectra were
individually corrected for interstellar extinction by Rose and Deng using the
reddening curve of Cardelli et al. (1989), as modified by O'Donnell (1994), 
with E(B-V) values from Peterson (1993).

\section{Stellar Populations in the Mid-UV}

\subsection{Choosing an Appropriate Mid-UV Color}

If we are to correctly interpret the absorption line measurements in the
mid-UV, it is important to compare these measurements to a UV measure of
temperature.  As shown in FOBW, the 2600--3000 color is a reasonable
compromise, in that: a) it can be self-consistently defined for mid-UV
spectra, as well as reliably measured; and b) it is sensitive to the stellar
population mix in the mid-UV.  Within that stellar population mix we cannot
ignore the contribution of the hot component which dominates the far-UV in
elliptical galaxies.  As shown in Burstein et al. (1988), this ``UVX''
component can contribute substantial flux at wavelengths as long as 3200 $\rm
\AA$ in some cases.  UVX effects are clearly present in our galaxy sample. 
They are seen, for instance, in the anticorrelation between the 2600--3000
colors and the optical Mg$_2$ indices for these galaxies (cf. the
data in Table~1 with those in Table~3a).

Tables~3a,b give reliable 2600--3000 colors for our program
objects.  Similarly, the stellar library from IUE (FOBW) has directly
measured values of 2600--3000.  Of the two other stellar libraries used here,
it is profitable to estimate 2600--3000 colors only for the line indices of
Davidge \& Clark (1994), as these are closest in wavelength to the mid-UV. We
list the measurements for additional indices from the Lick system in
Tables~3a,b for reference only. We display our results in the form of
(logarithmic) index-index diagrams below.

There are two main issues to be addressed here. First, do the galaxies and the
two globular cluster systems differ significantly from each other in any index
comparisons?  We will see that they do. Second, is it possible to synthesize
the composite systems (clusters or galaxies) by suitable combinations of the
Galactic stellar library objects?  We do not attempt actual spectral synthesis
modeling in this paper, so it is important to understand where potential
combinations would appear in the diagrams.  A given index formed from any
combination of stars is the weighted sum of the indices of its components,
where the weight $w_i$ for component $i$ is $0 \leq w_i \leq 1.0$ and $\Sigma
w_i = 1.0$.  However, the weights will change with wavelength.  This means that
unless the two indices involve a common normalizing wavelength, a combination
formed from two components will not fall exactly on the straight line
connecting them in a ratio-ratio diagram, and their relationship will be
further distorted in the logarithmic index diagrams shown here.

As discussed above, most of the galaxies have significant UVX contamination at
wavelengths below $\rm 3200 \AA$ but the M31 globular clusters do not. This
means that separations between clusters and galaxies in the index-index
diagrams could be the result of feature dilution or bluing caused by the UVX
component (which has a smooth energy distribution similar to that of an early
B-type star, with a 2600--3000 color of $\sim -0.5$). However, such
separations are not necessarily only the result of UVX contamination, and we
will try to distinguish between these two cases. On the other hand, those
diagnostic diagrams in which the galaxies and M31 clusters are superposed but
separated from the locus of Galactic library stars are indicative of other
explanations for the separation: e.g. differences in chemical abundances.

\subsection{Elliptical Galaxies and Globular Clusters in The Mid-UV}

For clarity, we compare the line indices and spectrum breaks determined in our
FOS G270 spectra and in the IUE LW spectrum for M32 to those for IUE-observed
stars in Figures 5a-g.  At the end of this section we will compare these
values with the IUE values for Galactic globular clusters. Those indices with
center wavelengths below 2550$\rm \AA$ were not used, as the IUE data are
generally too noisy shortward of this wavelength.  The data for galaxies are
plotted as open symbols; those for the M31 globular clusters are plotted as
closed symbols; the data for Galactic stars are plotted as small, subdued
symbols.  All error bars are smaller than the plotting symbol for the
2600--3000 color, save for NGC~5018, and are not shown.  Errors in the line
indices are explictly shown in Figures~5.  Each of the possible comparisons
among the galaxies, M31 globular clusters, Galactic globular clusters, and
Galactic stellar libraries is of interest.

Figure 5a: 2609/2660 break.  The M31 clusters follow the locus of Galactic
library stars well, but the galaxies other than M32 fall systematically
below, with 2609/2660 values at a given 2600-3000 color up to 0.4 mag
smaller.  This is probably a result of UVX contamination in the galaxies; the
implied contribution of the hot component in the galaxies at 2660 $\rm \AA$
is about 60\% in the most extreme case (NGC~7619; cf. \cite{BBBFL88}).
                                                                            
Figure 5b: 2828/2921 break. The values for the M31 globular clusters and the
galaxies agree well, and both lie at somewhat stronger index measures compared
to the stellar library at the same color.  The offsets from the Galactic
library locus here cannot be explained by UVX contamination in the galaxies.
The two weakest clusters (MII and MIV) have 2828/2921 values lower than those
of any galaxy.

Figure 5c: Mg II 2800 index.  This is the line index with the paradoxical
behavior with metallicity:  it gets weaker as metallicity increases. Moreover,
it is double-valued with 2600--3000 color (as is the 2609/2600 break), peaking
near $2600-3000 = 1.5$ (owing to chromospheric emission becoming greater with
later-type stars; cf. FOBW, \cite{smithetal}). Both galaxies and M31 globular
clusters show well-defined relationships between line index and color.  The
M31 clusters agree well with the locus of Galactic library stars.  However,
the galaxies have indices which are systematically $\sim 0.3$ mag smaller
than the clusters and the stars.  This could be another instance of UVX
contamination; however, for at least 3 of the galaxies the Mg I 2852 index
(see below) does not show a similar offset despite immediate proximity in
wavelength.  This argues against UVX contamination as the main cause of the
Mg II offset in the galaxies.

Figure 5d: Mg I 2852 index.  Despite involving the same element, lying only
50$\rm \AA$ from Mg II, and sharing a continuum sideband with the Mg II index,
in this Mg I index the M31 clusters and galaxies behave very differently than
in Mg II.  Indices for three of the clusters are significantly stronger than
in the Galactic library stars, and three of the galaxies are coincident with
these clusters. The other five objects fall close to the stellar locus.

Figure 5e: Fe I 3000 index.  As we have shown previously (FOBW), the Fe I 3000
index shows little direct correlation with metallicity among Galactic stars,
but does show a strong temperature dependence.  Here, five of the galaxies
and two of the clusters have indices significantly higher than the Galactic
library locus and the remaining four objects. This is the only UV index in
which the galaxies have conspicuously stronger index values than the M31
globular clusters (note that for most optical-band metallic features, gE
nuclei have stronger indices than do globular clusters).

Figure 5f: Blend 3096 index.  This index is a combination of Al I (at 3092.7
$\rm \AA$) and Fe I lines.  It was found in FOBW to have the clearest
luminosity segregation for late-type Galactic stars among the mid-UV indices.
Here, all of the M31 clusters and three of the galaxies have indices
significantly stronger than the Galactic library locus.  In contrast to Fe I
3000, the clusters form the upper envelope for the galaxies.  This is the
first index we have discussed where it is reasonably clear that the Galactic
library cannot be used to synthesize most of the enhanced cluster or 
galaxy indices and the anomaly cannot be attributed to UVX contamination
affecting the line index itself.  On the other hand, it is curious that the
three galaxies with the bluest measures of 2600--3000 (and, by inference, the
strongest UVX components) define Mg I 2852 and Blend 3096 relations with
2600--3000 colors that are {\it more similar} to those of the M31 globular
clusters than those defined by the other 4 galaxies.

Figure 5g: Mg Wide index.  We calculate this index as it is most likely the
index researchers will use to investigate the stellar populations of galaxies
at high redshift, when signal-to-noise and bandpass widths are limited. The
relationships defined by M31 globular clusters, galaxies and Galactic stars
are essentially coincident in this figure.

Comparison of the four Mid-UV line indices measured for Galactic globular
clusters by Rose \& Deng (1998) to those measured here for elliptical galaxies
and M31 globular clusters is shown in Figures 6a-d. Error bars for the HST
acquired data are supressed in this figure for clarity.  In all four diagrams,
the stellar populations of the Galactic globular clusters substantially overlap
those of the M31 globular clusters.  The offsets between M31 globular clusters
and galaxies in the 2609/2660 and Mg II 2800 indices are also seen between the
Galactic globular clusters and the galaxies. The Galactic globular clusters
define a range in Mg I 2852 vs. 2600--3000 that encompasses both M31 globular
clusters and all of the galaxies.  Note that two UV indices of M32 
(the galaxy lying at 2600--3000 = 1.50) have large differences from those of
the Galactic globular cluster 47 Tuc (2600--3000 color, Mg II 2800), to which 
it is often compared.  This reinforces the smaller amplitude distinctions at 
optical wavelengths discussed by Rose (1994).

\newpage

\section{The Near-UV and Blue Indices}

Line indices for features in integrated spectra in the near-UV, from 3100$\rm
\AA$ to 3600$\rm \AA$, just below the Balmer jump, are the current ``orphans''
of stellar population work.  In our perusal of the literature, we have only
found three studies that have explored this spectral line region in any depth
(\cite{DC94}, \cite{Carb82} (and associated references 1982-85), and
\cite{BRV88}). Of these, the line indices defined by Davidge \& Clark
are closest in spirit to the methodology employed here, while the
measurements of Carbon et al. and Boulade et al. differ in method and are
more specific to their own stellar population measurements.

As with the mid-UV indices, we would like to intercompare the near-UV
measurements of the elliptical galaxies and M31 globular clusters with those
of Galactic stars.  The current stellar data for all but the CN4170 feature
are effectively limited to the 16 stars observed by Davidge \& Clark. The
spectra of stars observed by Boulade et al. were not reduced to fluxes, and
hence their measures could not be used here without separate calibrations,
which are not available.  The CN4170 observations of Davidge \& Clark use the
Lick (\cite{worthey94}) bandpass definition for the ``CN1'' index. 74 stars
with Lick CN1 data were observed by FOBW, making this the only feature with a
large stellar sample of both near-UV feature data and mid-UV 2600--3000
colors. The 2600--3000 colors for the Davidge \& Clark stars are given in
Table~5a, while Table~5b lists the Worthey et al. (1994) CN4170 measures and
their 2600-3000 colors from Li et al (1998).

Four of the stars from Davidge \& Clark have published 2600--3000 colors in
FOBW, and three others have mid-UV colors from other IUE observations (as
maintained in our IUE Stellar Library (\cite{lietal98}, in preparation.). For
the remaining Davidge \& Clark stars we estimate their values of 2600--3000
from those given for the average stellar groups in Fanelli et al. (1992)
(using the colors given in Table~2 of that paper.  We note that owing to a
transcription error, the 2600--V and 3000--V colors for the group means listed
in Table 7B of \cite{FOBW92} are based on narrow-band rather than
intermediate-band fluxes, though the other tabulated quantities are correct.) 
These 2600--3000 colors should be accurate to 0.10 mag for the non-IUE stars,
based on making similar estimates for the Davidge \& Clark stars with measured
IUE colors.

Figures 7a-f compare the near-UV line indices for galaxies, M31 globular 
clusters and Galactic stars.

Figure 7a: NH 3360 index.  Given previous measurements of strong CN features
in the M31 globular clusters and early-type galaxies, measurement of this
index is central to the issue of possible nitrogen enhancement in these
systems.  The NH 3360 values for the Davidge \& Clark main sequence stars
earlier than K2~V are generally less than 3.00$\rm \AA$.  As shown in this
figure, the NH 3360 index is obviously far stronger in the M31 globular
clusters than in Galactic stars of similar 2600--3000 color.  While the
formal values of NH 3360 for the galaxies are stronger than for the Galactic
stars, the errors on these measures indicate that only the NH 3360 index
for NGC~7619 is signficantly above those for the stars, but weaker than those
for the M31 globular clusters.  These differences must be due to the strength 
of the NH 3360 index itself and cannot be due to UVX contamination.

Figure 7b: CN 3883.  The distribution of M31 globular clusters, galaxies and
stars is similar to that found for Bl 3580 and Fe I 3000, rather than as
found for NH 3360, in that this index seems to reach a maximum value. Since
this index is known to become saturated when the CN molecule becomes too
abundant (cf. \cite{langer}), it is not surprising that we see little
difference between the globular clusters and galaxies here, when we see more
of a difference between them in NH 3360.

Figure 7c. Ca HK.  This index is mildly enhanced in the more metal-rich
globular clusters and the galaxies, being of similar line strength for
both kinds of stellar populations at similar 2600--3000 colors.  The
degree of enhancement is small enough that it is plausible one might
be able to fit these measurments with a library of Galactic stars.

Figure 7d. BL 3580.  The CN feature at 3580$\rm \AA$ lies within this
passband (cf. \cite{Carb82}), but the feature seen is also a blend of many
lines due to mostly Fe-peak elements.  As such, the relative distribution of
values for the M31 globular clusters, galaxies and Galactic stars is very
similar to that found for the Fe I 3000 index in the mid-UV (Fig.~5e).  The
difference here is that most of the galaxies and globular clusters have
stronger Bl 3580, relative to the stars, than Fe I 3000.  That both galaxies
and globular clusters share this difference rules out the UVX component as
its main source. The fact that the strength of this blend is not one-to-one
with that of CN4170 (cf. Table~3A) for M31 globular clusters suggests that
CN is not the principle absorption source for Bl 3580.

Figure 7e. CN 4170.  The galaxies and M31 globular clusters define different
relationships between 2600--3000 and this CN index than they do with CN 3883
or NH 3360.  The globular clusters and three of the galaxies lie on a
well-defined locus offset by $\sim$0.15 mag to stronger CN from the galactic
stars.  Three galaxies have yet stronger CN at a given 2600--3000, but if
their colors were corrected for the UVX contribution, it is possible they
would lie on an extension of the M31 cluster locus.  The CN excesses for the
clusters and galaxies were expected on the basis of earlier studies
(e.g., \cite{BFGK84}; \cite{trip89}).

Figure 7f. Break 4000.  This break correlates reasonably well with 2600--3000
color, but shows little difference among M31 globular clusters, galaxies and
Galactic stars.  As such, this index is as disappointing as the Mg-wide index
(Fig.~5g), in that it does not distinguish differences in stellar populations
despite differences known in other Mg-- and N--related line indices.

Note that, in Figs. 7a-f, the line indices for the three galaxies with blue
2600--3000 colors (NGC~7619, 3608, 5831) show a relationship to those for the
redder 2600--3000 galaxies that is partly defined by this color difference. 
If we had used a color such as B--V for the horizontal axis of these graphs,
the line index--color relations would be continuous for the galaxies in these
diagrams.  However, the general relationship of the M31 globular clusters
relative to the galaxies would not be very different.

\section{Discussion}

It has been previously suggested on the basis of high precision optical and
infrared photometry or spectra (cf. \cite{frogel80}, \cite{O83},
\cite{BFGK84}, \cite{rose85}, \cite{Burstein87}, \cite{rose94}) that the cores
of elliptical galaxies, M31 globular clusters, and Galactic globular clusters
constitute three distinct types of stellar populations.  The mid-- and near--UV
observations discussed here substantially strengthen this view.  The more
subtle distinctions which are present in the optical bands are found to be
amplified in the mid-- and near--UV.  However, in the variety of index-color
behaviors exhibited, our data also emphasize the complexity of the situation. 
In this section we discuss these issues.

\subsection{2600--3000 Color vs. Optical Mg$_2$}

While the 1550--V color employed by Burstein et al. (1988) cleanly separates
out very hot stars in a stellar population, the 2600--3000 color depends on
the detailed mix of hot and warm stars. It is clear that for globular clusters
in our Galaxy and in M31, the dominant contribution of flux to the 2600--3000
spectral region comes from the brightest main sequence stars --- typically
main sequence turn-off (MSTO) stars.  The situation is more complicated for
galaxies, as the UVX component can contribute significant flux throughout this
wavelength region in the spectra of most giant ellipticals.

As shown in many studies (cf. \cite{frogel80}; \cite{BFGK84}) optical and
infrared broad-band colors of galaxies, M31 globular clusters and Galactic
globular clusters tend to track each other very well.  This is also true for
some optical absorption line indices, such as Mg$_2$, but less so for others
(such as the Lick iron line indices Fe 5270 and Fe 5335).  In the relationship
of UV/optical colors with optical line indices, Burstein et al. (1988) showed
that 1550--V color correlates well with Mg$_2$, albeit in the opposite sense
to optical colors (1500--V becomes bluer with stronger $\rm Mg_2$).

Figure~8 shows how $\rm Mg_2$ is related to 2600--3000 color for galaxies, M31
globular clusters and Galactic globular clusters.  To the data in the present
paper we have added the 2600--3000 colors for those Galactic globular clusters
with $\rm Mg_2$ measures by Burstein et al. 1984, as well as 2600--3000 colors
for representative galaxies from the study of Burstein et al. (1988), combined
with the new Trager et al. (1998) $\rm Mg_2$ measures.  We note that the 
aperture size difference between IUE ($\sim 8''$ effective diameter; cf.
Burstein et al.) and the Lick optical Mg$_2$ measures ($\sim 2''$ diameter)
did not significantly affect the 1500--V vs. Mg$_2$ analysis of Burstein et al.
Hence, we do not expect using IUE 2600--3000 colors to signficantly affect the
comparison we make with optical Mg$_2$ here.

It is evident that the 2600-3000 relation to Mg$_2$ for these stellar
populations is not monotonic.  M31 globular clusters and Galactic globular
clusters show a general increase in the strength of Mg$_2$ with redder
2600--3000 color but with the M31 clusters generally having stronger Mg$_2$
indices than Galactic clusters of similar 2600--3000 color {\it at all
colors}.  One galaxy in this sample, the S0 NGC~5102 (also NGC~205, not
shown), has colors similar to the bluest clusters, but this is a well-known
star-forming system in which the blue colors result from upper main sequence
stars (cf. discussion in Burstein et al. 1988); it is not representative of
the other objects in the sample.

The other galaxies appear to fall into two loosely-defined groups.  One group
of 4 (NGC~4382, NGC~5018, NGC~3605 and M32) is coincident with the reddest and
most metal-rich clusters.  A larger group has significantly higher Mg$_2$
showing relatively small scatter in the index but large 2600--3000 scatter. 
This is similar to what is seen when Mg$_2$ is plotted versus the 1550--V
color of Burstein et al. (1988), in that the line index varies little while
the 1550-V color varies over nearly 2 magnitudes.  This strongly suggests that
the UVX component that gives rise to the 1550--V range in color among
ellipticals is also important in determining the 2600--3000 range in color.

We can estimate how much UVX contamination would be necessary to produce the
blueward scatter of the strong-lined objects.  An extrapolation of the
relatively shallow slope of the line-color correlation in Figure~8 for
the clusters and the redder galaxy group predicts a 2600--3000 color
near 2.0 for objects with Mg$_2$ values comparable to those in the
strong-lined group.  Adopting a 2600--3000 color for the UVX objects
of $\sim -0.5$, we find that the UVX would have to contribute over
65\% of the light at 2600 $\rm \AA$ in order to shift the strong-lined
objects to 2600--3000 colors of $<$ 1.0.  This is reasonably consistent with
our estimate (\S 4) of over 50\% contamination of the UVX component of the 2600
bandpass when the UVX component is strong.

Obviously, a larger sample of UV data is needed to explore this correlation
further.  On the basis of the present sample alone, it appears that the
galaxies themselves may fall into several discrete groups.  Galaxies in several
of the other line index-color diagrams (notably BL 3096, Fig. 5f) also have a
clump-like distribution.  Hints of similar discreteness in the comparison of
fluxes at 1500\AA, 2500\AA, and the V-band with Mg$_2$ were pointed out by
Dorman et al. (1995). This would imply that metallicity, as measured by
Mg$_2$, is not the only driver of the amplitude of the UVX component.  A
similar conclusion is reached by Ohl et al. (1998) on the basis of UV color
and line strength gradients inside ellipticals.

As a separate issue, it is not clear why the M31 globular clusters should have
stronger $\rm Mg_2$ indices than Galactic globular clusters at a given
2600--3000 color.  As we are comparing HST FOS data with IUE data, it is
possible that systematic errors could play a role.  While reliable errors for
the IUE cluster data are difficult to determine, it is hard to see how the
errors could be as large as the 0.3--0.5 magnitude 2600--3000 color 
difference we do see.

\subsection{Nitrogen-Dependent Indices}

Burstein et al. (1981; 1984) first pointed out that CN line indices were
significantly stronger in M31 globular clusters than Galactic globular
clusters of the same B--V color.  This discovery was later confirmed by
Tripicco (1989).  As shown in Figure~7, the three near-UV line indices
involving nitrogen:  CN 4170, CN 3883 and NH 3360 show marked enhancements
relative to Galactic stars for both galaxies and M31 globular clusters. 
Whereas interpreting the strength of CN in terms of nitrogen abundance is
complicated by questions of carbon abundances and the mixture of main sequence
and giant-branch light (molecular abundances are very sensitive to surface
gravity), NH is notably cleaner because only the one metal is involved and the
light at 3300\AA\ is dominated by main sequence stars.
                                                  
The fact that the M31 globular clusters are over-enhanced the {\it most} in NH
3360 relative to Galactic stars is a clear indication that the M31 globular
cluster stars are overabundant in nitrogen.  How overabundant? Look at the
spectrum of MIV, the globular cluster in M31 with similar colors and optical
line strengths as M92 (NGC~6341), one of the most metal-weak globular clusters
in our own Galaxy.  Carbon et al. (1982) studied the temperature, gravity and
abundance dependence of NH 3360 in the spectra of the giant stars of M92. They
show that, in this low abundance range, the NH 3360 feature is little
dependent on gravity, but primarily dependent on temperature and abundance. In
particular, NH 3360 all but disappears in stars with effective temperatures
hotter than 5500 K.  The MSTO stars in M92 are hotter than this temperature,
and it is these stars that contribute the bulk of the flux at 3360$\rm \AA$ in
this metal--poor cluster (cf. discussion by \cite{rose94}), and, by
association, in MIV.  

Moreover, it seems that the NH feature reaches something approaching a
``saturation level'' in the more metal-rich M31 globular clusters, reaching a
near-constant value of $\rm \sim 7.5 \AA$.  Such an effect is also seen in the
CN~3883 index where it is known to be a saturation effect (\cite{langer}). The
strong enhancement of both NH~3360 and CN~4170 in M31 globular clusters
relative to Galactic stars means that the enhancement of CN~4170 in M31
globular clusters relative to Galactic globular clusters found by Burstein et
al. (1984) is also due to a nitrogen overenhancement. Significantly, the 
NH~3360 index for these galaxies is also enhanced relative to Galactic stars, 
but is generally lower than for the M31 globular clusters.  (Interestingly,
Boulade et al. (1993) found NH3360 strength in M32 to be reasonably fit with a
Galactic stellar population.)  Thus, we are left with another unanswered
question:  Why are many old stellar populations outside our Galaxy enhanced in
nitrogen abundance relative to the disk and halo stars near us and to the
stars in Galactic globular clusters?

\subsection{Mg-related Indices}

It has already been shown by Gonzalez (1993) and Trager (1996) that the
optical $\rm Mg_2$ index is overly strong in the stellar populations of
early-type galaxies, compared to optical Fe-based indices and stellar
population models made from Galactic stars.  The M31 clusters also appear to
be significantly stronger in Mg$_2$ than Galactic clusters, compared to their
2600--3000 colors, though both are significantly weaker in Mg$_2$ than most 
galaxies (Fig.~8; see, however, the Mg$_2$ - B--V relations for these same
clusters in \cite{BFGK84}).

Paradoxically, these relationships fail to hold for the UV Mg I and Mg II
indices, relative to the 2600--3000 color.  Behavior of the Mg I 2852 index
(Fig.~6d) for the two sets of clusters may parallel that of Mg$_2$ in Fig.~8,
but the two cluster samples appear to overlap in the Mg II 2800 index 
(Fig.~6c). Furthermore, the galaxies are no stronger in Mg I 2852 than the 
clusters and are actually significantly weaker in Mg II.

Given the optical-band Mg$_2$ results, the galaxies are clearly not
underabundant in magnesium itself.  There are at least three possible effects
which may operate to produce the UV Mg anomalies in galaxies. First is UVX
contamination, which could dilute the Mg II strengths, though as noted in
Sec.~4.2 this does not seem capable of explaining the entire Mg II anomaly. 
Second is the fact that the Mg II index actually declines in more metal-rich
stars, apparently owing to large line blanketing in the sidebands (FOBW,
\cite{FOBW92}).  The third possibile effect is Mg II line reversal by
chromospheric emission.  This is clearly detectable in individual stars
(\cite{smithetal}). However it is only present in stars younger than 1--2 Gyr
(according to conventional Ca II chromospheric emission dating scales).  While
there is good evidence for the presence of intermediate age populations in the
range 4--8 Gyr in some of our sample elliptical galaxies (Gonzalez 1993,
Trager 1996), components with ages in the 1--2 Gyr range are usually
excludable because of their gross effect on the continuum energy distribution.
However, it may be that the chromospheric and galaxy spectral age scales are
not yet mutually consistent.

As a separate consideration, it is clear from the distinct behavior of the
elliptical galaxies and star clusters in the Mg indices that a metal-poor
(cluster-like) stellar population is {\it not} likely to be dominant at
far--UV wavelengths in the galaxies, as had been claimed by Park \& Lee
(1997), among others.

\subsection{The Other UV Indices}

\subsubsection{Breaks 2609/2660 and 2828/2921}

The 2609/2660 index is the most affected by flux from the far-UV strong
stellar population in galaxies among all of the indices measured in this
paper. The galaxies as a group are offset from the globular clusters as a
group by $\approx 0.3$ mag.  As we know the ellipticals have a strong UVX
component in their UV spectra that is missing in the globular clusters, it is
likely that the index offset is explained mainly by UVX contamination. Rose \&
Deng (1998) also show that the offset between 47 Tuc and M32 (the reddest
galaxy in Fig.~6a) can be explained by the weak UVX component (\cite{BBBFL88})
in M32 that is not in 47 Tuc.

The 2828/2921 index shows reasonable agreement among all of the globular
clusters and galaxies, with perhaps an intrinsic spread of about 0.1 mag. 
Given this general agreement, it is likely that this index can be fit in all
of the stellar populations with the normal Galactic library.  UVX
contamination is evidently considerably less at 2900\AA\ than at 2600\AA.

\subsubsection{Fe I 3000 and Bl 3580}

Both of these indices are dominated by a series of strong Fe I lines,
although the BL 3580 index is also somewhat affected by CN. The apparent
excess line strengths in Figs. 5e and 7d may be explainable by compositeness
effects --- i.e., the combination of cool and warm stars from the normal
Galactic library, together with the presence of hot UVX stars in some of the
galaxies.  Detailed modeling would be required to confirm that, however.

\subsubsection{Blend 3096}

This is the first time this feature has been reliably measured in external
stellar populations.  Examination of the solar spectrum (\cite{moore66}) shows
the strongest lines in the central bandpass of 3086--3106$\rm \AA$ are due to
Al I (at 3092.7$\rm \AA$) and Mg I (at 3096.9$\rm \AA$), combined with weaker
lines of Fe I, Ti II, Ni I and OH.  To our surprise, this index is so strong
in the strongest-lined globular clusters and in NGCs~3608, 5831 and 7619
that it appears no combination of Galactic stars can fit it.

A close up examination of the observed spectra of the M31 globular clusters
(where internal velocity dispersion does not erase fine detail) shows this 
feature to be double-peaked in absorption, with the separation
of the two peaks of about 6$\rm \AA$.  One peak is around 3092$\rm \AA$, and
can be reasonably associated with the Al I line.  The other peak is around
3098$\rm \AA$, and is likely a combination of the strong Mg I feature plus
the relatively strong Fe I lines between 3099--3101$\rm \AA$.

The strength of this feature in the M31 globular cluster spectra and in 
a subset of the galaxies is such that it is not just due to the known
Mg I enhancement, but also must involve an enhancement in aluminum as well.
This is the first time, to our knowledge, that an enhancement in Al has
been found in extragalactic stellar systems.

\subsubsection{Ca HK, 4000$\rm \AA$ Break, Mg Wide}

These three low-resolution line indices measure some of the most prominent
features in the integrated spectra of old stellar systems:  strong breaks in
the spectra around the Ca H\&K features near 4000$\rm \AA$ and near the Mg I
2852 and Mg II 2800 features.  These are the line strengths workers have used
to try to measure evolution in stellar populations at high redshift (e.g.
\cite{wind94}).  It is therefore disappointing that, whereas we can identify
three separate stellar populations among M31 globular clusters, elliptical
galaxies and Galactic stars/globular clusters in other line indices, these
three indices are insensitive to these distinctions. As Rose (1994) and others
have pointed out, lower resolution measures of old stellar populations can 
suggest a deceptive uniformity.

\subsection{The M31 Globular Cluster HST Color-Mag Diagrams}

Fusi-Pecci et al. (1996) have published color-magnitude diagrams (V vs. B--V
and I vs. V--I) for the four M31 globular clusters in our study.  While the
metallicities estimated from the giant branches for MIV, MII and K280 are
reasonably in agreement with the metalliticies we would estimate from our FOS
observations, K58 stands out as a notable exception.  Whereas Fusi-Pecci et
al. would have K58 be only slightly more metal-rich than MII, both the Brodie
\& Huchra measurement of its optical Mg$_2$ index, as well as our FOS spectra,
indicate this globular cluster likely has somewhat above solar metallicity. 
Some confusion in ranking the clusters via the CM diagrams might be the result
of using V--I for some and B--V for others (e.g., K58).  A V--I diagram for
K58 would be interesting, as it is possibly metal-rich enough to have its I
band flux saturated via an extreme nitrogen overabundance, much like Frogel et
al. result for the bulge of our Galaxy (\cite{frogel87})?  Clearly accurate
integrated spectra of these, and other M31 globular clusters into the
near-infrared would be of interest.

\section{Conclusions}

The same two elemental "suspects" come up time and again when intrinsic
differences among stellar populations are discussed: nitrogen and magnesium.
To this list, we now can perhaps add aluminum based on the behavior of the 
Bl~3096 index. One important contribution of our observations is that variations
in nitrogen abundance have been cleanly identified. It is clear from the
enhanced NH~3360 features, especially in the M31 globular clusters,
that as a group these four clusters have stars that are substantially
enhanced in nitrogen compared to Galactic stars. Based on the models of Carbon
et al., it is possible that the overabundance reaches factors of 10 -- 50 in
the M31 clusters. Nitrogen abundance estimates based on the CN indices
longward of 3800\AA\ are made ambiguous because of the involvement of carbon
and because the indices are more sensitive to the mixture of giant and dwarf
light. The value of the NH~3360 feature in old stellar population spectra is
that it isolates nitrogen and that dwarf light dominates at this wavelength.

The inferred large overabundance of nitrogen in the M31 globular clusters does
not conflict with the known agreement of broad-band colors between M31 and
Galactic globular clusters.  Nor does it conflict with the fact that the
HST-derived color-mag diagrams of these same globular clusters (cf.
\cite{fusipec96}) yield giant branch slopes generally consistent with their 
derived broad-band colors.  As Carbon et al. show, stars with very different
atmospheric abundances of nitrogen exist side-by-side in the color-mag diagram
of M92.  Moreover, theoretical studies by VandenBerg (e.g. \cite{vdbS}) show
that increasing only the oxygen abundance does not change the structure of the
HR diagram, but rather makes the cluster look older than it really is.  As the
same is likely to be true for nitrogen, HR diagrams alone will not divulge
nitrogen abundance differences among different stellar populations.

The combination of enhancements in N, Mg and perhaps Al are unusual, as each
is thought to come from different kinds of stars:  N from low mass stars; Mg
direct from C-- and O-- burning; Al from a combination of C-- and Ne--
burning, but also tied directly to O abundance. Laird (1985) and Boulade et
al. (1988) both found low metallicity stars with large (5--50) overabundances
in nitrogen.  Norris \& Smith (1983) point to a possible connection between
overenhancements of N, Na and Al in Galactic globular cluster giant stars. 
It would obviously be useful to attempt to measure Na abundances in
elliptical galaxies from prominent features like the Na I D lines at high
enough spectral resolution that contamination from interstellar absorption can
be eliminated.

Our observations have also emphasized the importance of the effects of the hot
UVX component at wavelengths as long as 3000\AA.  This is a potentially
serious complication for the age-dating of high redshift galaxies based on
restframe mid--UV spectra.  At present, we do not understand the UVX component
well enough to predict its strength from first principles; only direct
observations of the restframe far--UV spectra suffice to determine its
amplitude.  This is particularly true if it is not simply related to a single
parameter like metallicity, as suggested by the apparent discreteness of the
galaxy distribution in some of our correlations.

Other implications of these results for the study of stellar populations of
early-type galaxies, both nearby and at high redshift, will be discussed
further in Ponder et al. (1998).

\section{Acknowledgements}

     DB, CCW, RWO, JAF and JAR all wish to thank and acknowledge the support
of NASA Grants GO-2298-87A and GO-6585-95A for this research, and the referee
for helpful comments. DB wishes to thank the hospitality of the Space
Telescope Science Institute during his visits and the assistance of Mr. Yong
Li.  CCW specifically wishes to acknowledge that this project is supported by
the research contract GO-06585.04-95A awarded to the Computer Sciences
Corporation by Space Telescope Science Institute which is operated by the
Association of Universities for Research in Astronomy, Inc., for NASA under
contract NAS5-26555.  RWO was also supported in part by NASA grant NAG5-6403
to the University of Virginia.


\appendix

\section{Appendix A. Target Acquisition}

Precise centers of diffuse objects such as galaxies and globular clusters are
difficult to determine from ground-based data.  As such, the FOS must employ a
search procedure to place the center of these objects in the science aperture.
In acquiring each object, the telescope scans an area of sky around the input
coordinates. The target acquisitions use various apertures and scanning
patterns to determine the position with the greatest flux for that aperature.

The first (3$\times$1) scan ($4.3\arcsec$ aperture) for each object consisted
of three exposures. The second (2$\times$6) scan uses the $1.0\arcsec$
aperture and sweeps out a rectangle of six by two integration positions.  The
third (3$\times$3) scan takes an s-shaped pattern of integrations with the
$0.5\arcsec$ aperture.  The fourth scan (4$\times$4) also sweeps out an
s-shaped pattern with a $0.3\arcsec$ aperture.  Integration times per position
in these scans are necessarily short (5s--60s) to keep the acquisition times
within reasons (even then, they typically take 2-3 orbits). The fact that the
M31 globular clusters have small apparent effective radii, combined with
relatively accurate initial positions, allowed us to point to these objects
effectively with only the first three acquisition scans.  In the case of the
elliptical galaxies, the uncertainity of the positions of the cores of the
galaxies prompted us to use all four sets of acquisition scans.

As the FOS could saturate on these aquisition scans, a reasonable estimate of
expected flux was required to maximize the set-up signal while not getting too
much flux.  As we knew best the UV fluxes of our sources from their IUE
observations, it was decided to use the G270H grating as part of the
acquisition procedure.  This had the added benefit of being able to see in
each acquisition scan the spectrum of the position sampled.  To verify the
accuracy of the target acquisition, we reduced each of these scanned spectra,
and plotted them using the celestial positions from their header files.  An
example of one 3$\times$3 scan of one M31 globular cluster (K58) is shown in
Figure~9 as an example.  The S/N seen in these aquisition spectra are poor, in
keeping with the short exposure times. Due to the changing sizes of the scan
apertures, the positions of greatest flux in each successive set of spectra
can shift within one scan position, owing to the quantized scan search.

For each scanned spectrum, we determined its 2600--3000 color for both
intercomparison and comparison to the color we obtained with our program
spectrum.  By comparing the color for each scanned spectrum, the determined
centers, and the observed spectrum, we can determine the reliability of each
acquisition.  The colors determined from the science observations of K280,
MII, and MIV agree with the centers determined by the 3$\times$3 scan.  As
seen in Figure~10, the agreement is worse for the globular cluster K58, which
has the lowest UV surface brightness among the four M31 clusters.  Even so,
the aquisition colors are within nominal expected limits of the actual 
observed 2600-3000 color as given in Table~3A (2$\sigma$ for this many
observations and the poor S/N of the acquistion spectra).

For the elliptical galaxies, obtained in Cycles 4 and 6, the FOS aquisition
procedure used all four sets of scans to acquire diffuse objects.  In every
case, we confirm accurate registration of the FOS observation aperture with
the source of maximum UV flux in the galaxies.

\newpage

\begin{figure}
\figurenum{1a}
\caption{The de-redshifted, extinction-corrected FOS G270H spectra of the four
M31 globular clusters.  The spectral range is 2250 \AA \  to 3270 \AA. The
identification of the mid-UV absorption and spectral break features 
used in this paper are indicated on the spectrum of MII.}
\end{figure}

\begin{figure}
\figurenum{1b}
\caption{Same as Figure 1a, but for the FOS G400H grating --- a spectral range 
of 3200 \AA \ to 4780 \AA.  The identification of near-UV line absorption 
and spectral break features are indicated on the spectrum of MII.}
\end{figure}

\begin{figure}
\figurenum{2a}
\caption{The de-redshifted, extinction-corrected FOS G270H spectra of the six
elliptical galaxies. The spectral range is 2250 \AA \ to 3270 \AA.}
\end{figure}

\begin{figure}
\figurenum{2b}
\caption{Same as Figure 2a, but for the FOS G400H grating --- a spectral range 
of 3200 \AA \ to 4780 \AA.}
\end{figure}

\begin{figure}
\figurenum{3}
\caption{These figures show the full FOS spectra of the M31 globular clusters
given on a logarithmic flux scale from 2250 \AA to 4780 \AA.  The FOS G270H
spectrum overlaps that of the G400H spectrum by 70 \AA.  As can be seen, the
absolute flux levels of the two gratings agree to within 5\%.}
\end{figure}

\begin{figure}
\figurenum{4a}
\caption{The full FOS spectra of three of the elliptical galaxies, NGC~3605,
NGC~3608 and NGC~5018, given on a logarithmic flux scale from 2250 \AA to 4780
\AA.  The agreement in overall flux level between the G270H and G400H spectra
is well within 5\%, and is better than for the spectra of the M31 globular
clusters.}
\end{figure}

\begin{figure}
\figurenum{4b}
\caption{The full FOS spectra of three of the elliptical galaxies, NGC~5831,
NGC~6127 and NGC~7619, given on a logarithmic flux scale from 2250 $\rm \AA$ 
to 4780 $\rm \AA$.  Also plotted is the IUE merged LWP spectrum for M32, 
as taken from Buson et al. (1990).}
\end{figure}

\begin{figure}
\figurenum{5a,b}
\caption{This figure is in 7 parts (a--g), showing how the mid-UV line indices
and spectral breaks derived from the FOS spectra of the four M31 globular
clusters, the FOS spectra of the six elliptical galaxies, and the IUE spectrum
of M32 compare to those derived for Galactic stars from IUE LWP spectra by
FOBW.  In each case, the X-axis is the mid-UV color 2600--3000 (in magnitudes).
All line indices and spectral breaks are also measured in magnitudes. In each
figure the small, subdued symbols indicate the stars taken from FOBW.  Only
dwarfs (open squares), subgiants (open triangles) and giants (open diamonds)
are plotted.  The large closed symbols are the data for the M31 globular
clusters; K58 (square), K280 (triangle), MII (inverted triangle), and MIV
(diamond).  The large open symbols denote the galaxies: M32 (circle), NGC 3605
(inverted triangle), NGC 3608 (asterisk), NGC 5018 (diamond), NGC 5831
(triangle), NGC~6127 (square), NGC 7619 (cross).  Only the 2600--3000 error
bar is given for NGC~5018; for all other HST-observed objects the 2600--3000 
error is smaller than the plot symbol size.  (a) The 2609/2660 spectral break. 
(b). The 2828/2921 spectral break.}
\end{figure}

\begin{figure}
\figurenum{5c,d}
\caption{Same symbols as for Figures 5a,b. c) The Mg II 2800 line index.
d) The Mg I 2852 line index.}
\end{figure}

\begin{figure}
\figurenum{5e,f}
\caption{Same symbols as for Figures 5a,b. e) Fe I 3000 line index.
f) The BL 3096 line index, mostly measuring Al, Mg and Fe.}
\end{figure}

\begin{figure}
\figurenum{5g}
\caption{Same symbols as for Figures 5a,b. g) The Mg Wide line index, 
basically measuring the break in the continuum produced by the 
Mg II 2800 and Mg I 2852 lines.}
\end{figure}

\begin{figure}
\figurenum{6}
\caption{This figure redisplays Figures 5a--d, but this time adding the
Galactic globular clusters from Rose \& Deng (1998) and not plotting the data
for Galactic stars.  Error bars are supressed for clarity.
In this way, the comparisons among the Galactic globular
clusters, M31 globular clusters and elliptical galaxies are more clearly seen.
Data for Galactic globular clusters are plotted as small closed squares, for
M31 globular clusters as large open circles, and for galaxies as large open
triangles. (a) 2609/2660 spectral break. (b) 2828/2921 spectral break. 
(c) Mg II 2800. (d) Mg I 2852.}
\end{figure}

\begin{figure}
\figurenum{7}
\caption{The near-UV absorption line indices and the 4000 $\rm \AA$  spectral
break taken from the G400H FOS spectra.  Small symbols here represent the data
for Galactic stars taken from the study of Davidge \& Clark (1994) for all six
indices, supplemented by data taken from the stars given in Worthey et al.
(1994) for the CN4170 index.  The large symbols represent the four M31
globular clusters and the six galaxies observed with the FOS G400H grating,
using the same coding as in Figures 5a-g.  Errors in 2600--3000 are smaller
than symbol size for the HST data in these figures, and are not shown.
a) NH 3360; b) CN 3883; c) Ca HK; d) Bl 3580; e) CN 4170; and 
f) 4000 $\rm \AA$ spectral break.}
\end{figure}

\begin{figure}
\figurenum{8}
\caption{The optical line index, Mg$_2$ plotted versus the 2600--3000 color
for four sets of data:  M31 globular clusters (closed circles); Galactic
globular clusters (open squares); the six elliptical galaxies studied here
(closed triangles) and a representative set of early-type galaxies taken from
the IUE galaxy study of Burstein et al. (1988) (open triangles). Note the
range of Mg$_2$ at a fixed 2600--3000 color, indicating the relationships of 
optical colors and line indices with 2600--3000 color are complicated.}
\end{figure}

\begin{figure}
\figurenum{9}
\caption{The three by three G270H spectral scans of the M31 globular cluster
K58 taken during target acquisition with the FOS.  This is by far the worst
case among the 10 FOS object acquisitions done for this project. The box size
for each spectrum does not represent the size of the aperture (for $3\times3$
scanning, aperature size is $0.5\arcsec$).  The small circled x in Box 7
indicates that this was the position that the acquisition process picked for
maximium flux.  Comparing to Box 6, we see indication that there may be a
misalignment. But, because the aperture sizes are in general larger than the
amount by which the telescope has been shifted, the position of greatest flux
may not appear to be the optimium position.}
\end{figure}

\begin{figure}
\figurenum{10}
\caption{How well did the FOS aquisition work?  This figure plots the
2600--3000 colors derived from each acquistion G270H spectrum.  The symbols
used represent the colors obtained from the individual scans (small symbols:
open square for the initial 3$\times$1 scans with the 4.3$''$ full aperture;
open diamonds for second 2$\times$6 scans with the 1.0$''$ aperture; open
triangles for the third and final 3$\times$3 scans with the 0.5$''$ circular
aperture). The color for the scan center selected by the aquisition process
for each scan is denoted by a large symbol (cross for 3$\times$1 center; plus
sign for 2$\times$6 center; circle for 3$\times$3 center).  The 2600-3000
color obtained from our program observation is indicated by a large square.
Note that for all but K58, the 2600-3000 color from the 3$\times$3 center and
that from the program observation are very similar.  For K58 the difference is
greater, owing to the lower S/N spectra for this cluster.}
\end{figure}

\end{document}